# Satellite reveals age and extent of oil palm plantations in Southeast Asia

Olha Danylo[1], Johannes Pirker[1,2], Guido Lemoine[3], Guido Ceccherini[3], Linda See[1], Ian McCallum[1], Hadi[1], Florian Kraxner[1], Frédéric Achard[3], Steffen Fritz[1]

[1]Ecosystems Services and Management Program, International Institute for Applied Systems Analysis, Schlossplatz 1, A-2361, Laxenburg, Austria.

[2]KU Leuven, Department of Earth and Environmental Sciences, Leuven, Belgium

[3]Institute for Environment and Sustainability, European Commission, Joint Research Centre, I-21020 Ispra (VA), Italy

**Satellite reveals age and extent of oil palm plantations in Southeast Asia**

**Summary**

In recent decades, global oil palm production has shown an abrupt increase, with almost 90% produced in Southeast Asia alone. Monitoring oil palm is largely based on national surveys and inventories or one-off mapping studies. However, they do not provide detailed spatial extent or timely updates and trends in oil palm expansion or age. Palm oil yields vary significantly with plantation age, which is critical for landscape-level planning. Here we show the extent and age of oil palm plantations for the year 2017 across Southeast Asia using remote sensing. Satellites reveal a total of 11.66 (± 2.10) million hectares (Mha) of plantations with more than 45% located in Sumatra. Plantation age varies from ~7 years in Kalimantan to ~13 in Insular Malaysia. More than half the plantations on Kalimantan are young (<7 years) and not yet in full production compared to Insular Malaysia where 45% of plantations are older than 15 years, with declining yields. For the first time, these results provide a consistent, independent, and transparent record of oil palm plantation extent and age structure, which are complementary to national statistics.

**Main body**

Global oil palm (*Elaeis guineensis*) production has more than doubled over the last two decades driven by food and industrial demand. Plantation area went from 10 to 21 Million hectares (Mha), while oil production increased from 100 to 300 million tons (Mt) crude oil palm between 1997 and 2017. Almost 90% of the cultivated oil palm is located in Southeast Asia[1]. This surge has increased public revenues and helped to alleviate rural poverty for millions of farmers and agricultural workers[2,3], but it has also deeply impacted natural forest ecosystems[4–7] and biodiversity[8,9], while contributing significantly to climate change by releasing carbon from converted forests and peatlands into the atmosphere[4,10,11]. Based on the land sparing theory[12,13], the improvement of oil yields has been put forward as a means of reconciling oil production and forest conservation[14,15]. Oil palm is known to be the most efficient oil producing plant globally. Oil palm yields, however, vary dynamically with plantation age. They increase during the youth phase of the first seven years, reach a plateau during the prime age of 7-15 years, and then slowly start to decline before palms are replaced at the age of 25-30 years[16]. Therefore, knowing the exact extent and age of plantations across a landscape is crucial for landscape-level planning to allow for both sustainable oil palm production and forest conservation.

Today, the combination of high-resolution satellite records and cloud-computing infrastructures to handle big data can provide a complementary asset to quantify oil palm extent and age that is independent of official statistics. The latter are usually provided annually at a rather coarse spatial scale (national or regional administrative units), they may be incomplete or not updated regularly, and the quality is uncertain. Hence, they do not provide either detailed spatial extent nor near-real time trends in oil palm expansion or age.

Several recent studies have applied remote sensing techniques to estimate the extent of oil palm plantations in Southeast Asia. Studies aimed at quantifying plantation extent at a national scale or larger, have generally deployed visual, expert-based interpretation methods[4,17–20] or semi-automatic approaches, blended with extensive field information[21,22]. These approaches are, however, very labor intensive, which limits their utility for upscaling and developing a near real-time monitoring system. Some studies have employed radar imagery, which have benefitted from penetration through clouds, improved resolution and high revisiting frequency[23–25]. Recent experimental studies have demonstrated the usability of free and open Sentinel data to detect oil palm plantations[26]. Sentinel data have proved particularly useful in detecting smallholder plantations[27] and plantation age[28], which in turn, is a good predictor of oil palm yields[28,29].

Here we estimate the spatial extent of oil palm for Southeast Asia using radar and optical imagery from the Sentinel data set using Google Earth Engine, a big data Earth Observation platform that allows seamless parallel computing and geospatial operations (more details can be found in the Methods section). Our methodology has produced the first map of both industrial (i.e., large-scale, intensively managed) and smallholder oil palm plantations (i.e., smaller in size and generally reliant on family labor) at a 10 m resolution for Indonesia, Malaysia and Thailand for 2017. We have built on the large body of literature regarding the use of satellite remote sensing: 1) to deal with the specificity of oil palm cultivars (different management types, etc.) and 2) to propose innovative tools to support monitoring of oil palm plantations. The map accuracy ranges from 80% for Thailand to 85% for Peninsular Malaysia, with no clear trend in commission or omission of reference oil palm plantations. Comparison with other oil palm mapping products[18,24] demonstrates that the present oil palm map is more accurate for most regions. We also determine the year of plantation establishment back as far as 1984, producing remotely-sensed statistics of plantation age distribution. This is the first study to estimate spatially-explicit age classes of oil palm in addition to spatial extent. We performed cross-validation of plantation age by ensuring consistent reporting of the timing of forest disturbance by comparing age with an independent forest loss data set[30] as well as national agricultural statistics (see Methods for more details). The results are presented by major producing regions: (i) Sumatra and Kalimantan in Indonesia, (ii) Insular and Peninsular Malaysia and (iii) southern Thailand.

This evidence-driven assessment targets three key questions: i) what is the current spatial extent of oil palm in Southeast Asia? ii) following the recent boost in the market demand, are oil palm plantations changing throughout Southeast Asia, and if so, in which countries and to what degree? and iii) what is the age structure of oil palm plantations in Southeast Asia and how have these changed over time?

**Current spatial extent of oil palm in Southeast Asia**

In 2017, oil palm plantations covered an area of 5.22 Mha (±0.88 Mha) or 11% of the island's land surface on Sumatra, a further 2.50 Mha (±0.50 Mha) on Kalimantan, 2.11 Mha (±0.31 Mha) on Peninsular Malaysia, 0.84 Mha (±0.14 Mha) on Insular Malaysia, encompassing Sabah and Sarawak, and 0.62 Mha (±0.12 Mha) in Thailand. Two types of plantation are present. The first are industrial oil palm plantations, which dominate many regions and are characterized by large, structured and efficient patterns of contiguous blocks (dominant in Figures 1c to d), which makes them easily detectable to the human eye[18]. The second are smallholder plantations (prevalent in Figure 1a), which are reported in the official statistics[31], making up around 40% of oil palm production in Indonesia and as much as 70% in Thailand[32]. Smallholder plantations tend to be smaller, less structured and more heterogeneous in space and age structure and are hence more difficult to detect. In many landscapes, industrial and smallholder plantations co-exist next to each other, resulting in unspecific patterns of plantations of various sizes (see Figure 1b). While our method can identify smallholder plantations, it does not formally distinguish them from industrial plantations.

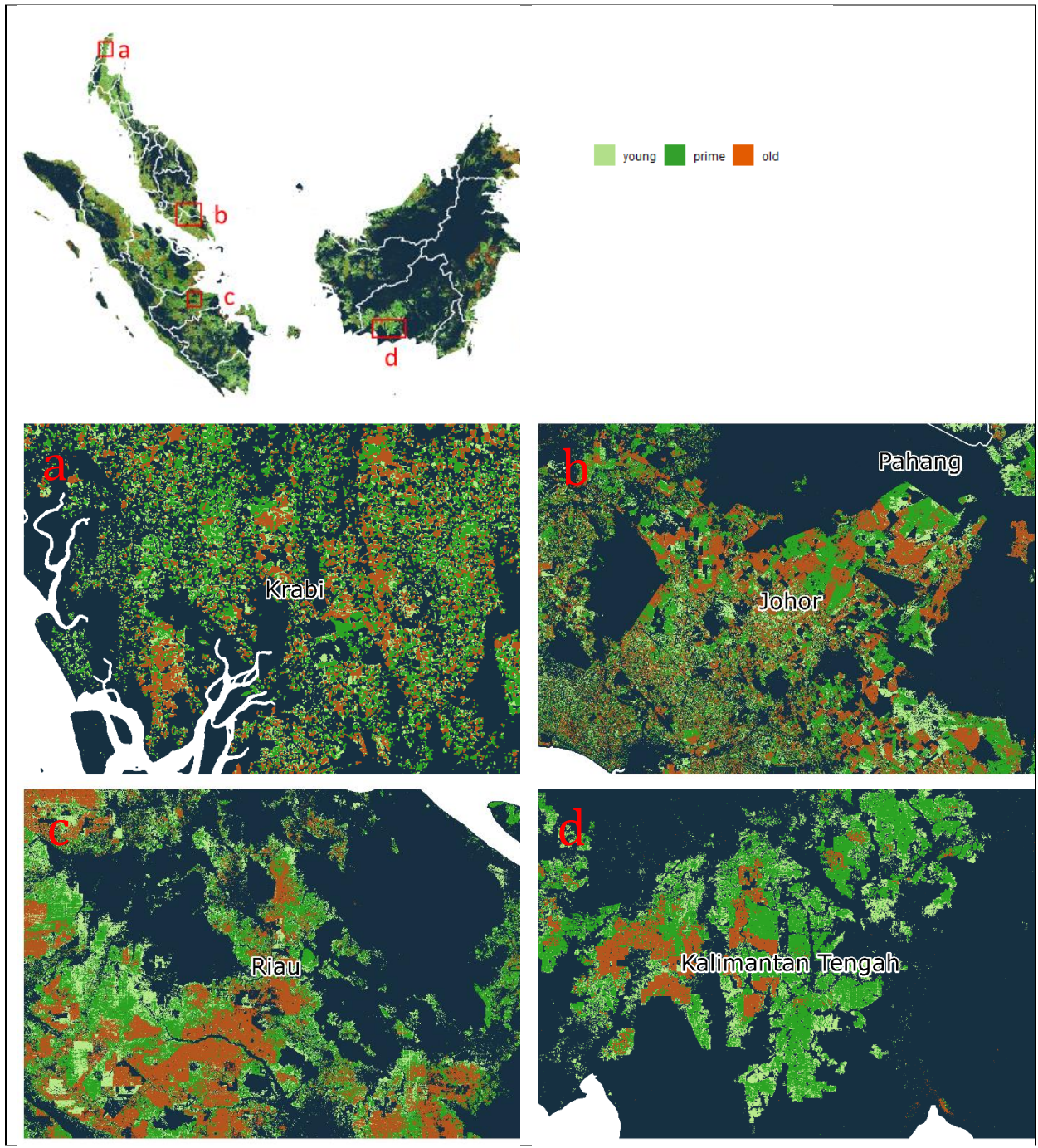

**Figure 1. Extent of oil palm plantations.** Maps of extent of oil palm plantations (by functional age classes: young <7 years; prime 7-15; old >15 years) across Southeast Asia (top left) and zooming into four selected locations in (a) Krabi/Thailand, (b) Johor/Peninsular Malaysia, (c) the province Riau in Sumatra, and (d) Southern Kalimantan.

**The temporal dynamics of oil palm plantations in Southeast Asia**
Figure 2 shows the expansion of oil palm plantations across Southeast Asia since 1984. The oil palm frontier first advanced in Malaysia and the Indonesian island of Sumatra, which soon became a major production area with ~50,000 ha of new plantations established per year around 1990, culminating in an all-year high of 380,000 ha of new plantations in 2012. On Kalimantan, large-scale expansion only started in the late 1990s, and around 2010, it became a major producing area. Expansion is still on the rise in the most recent years of the analysis, and 2015 saw the addition of 293,500 ha of new plantations – the single biggest conversion event in this region. In Thailand, oil palm plantations are expanding at a lower average rate of 35,000 ha per year.

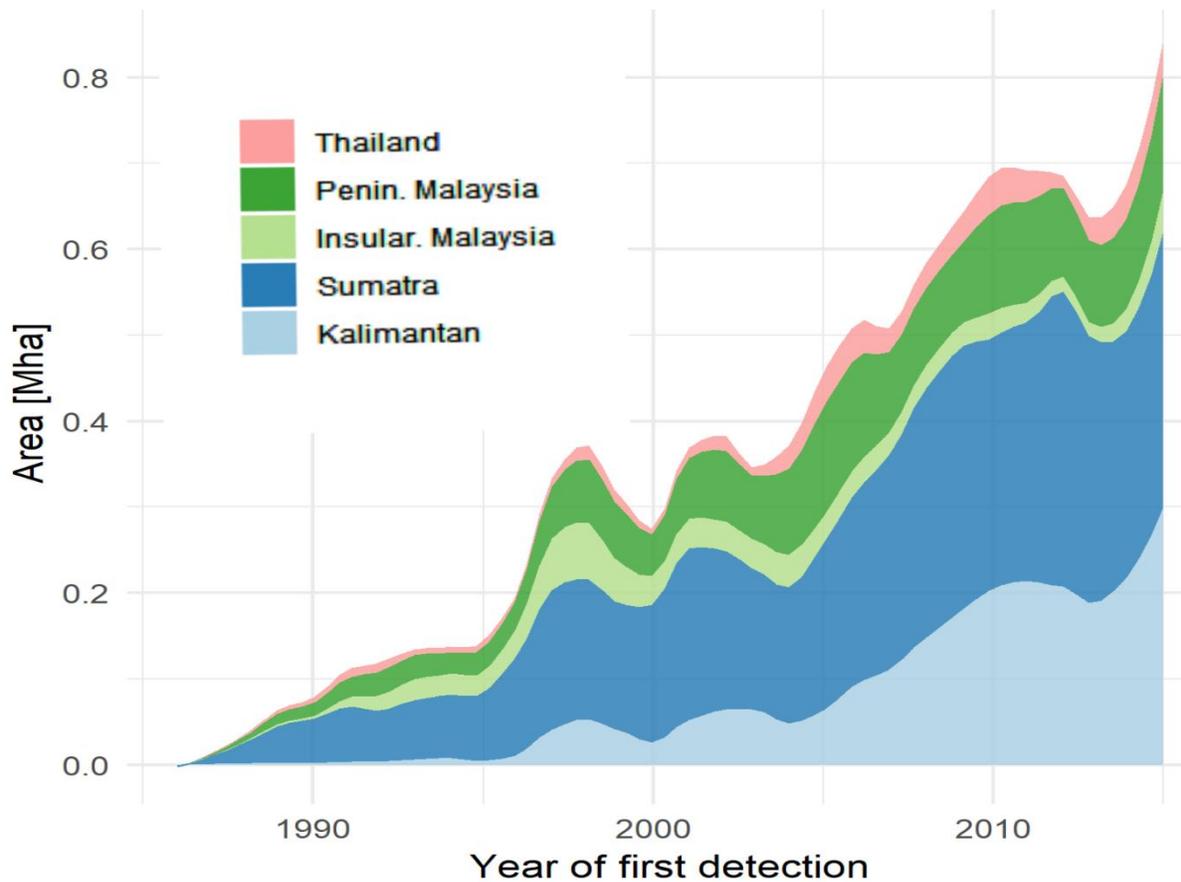

**Figure 2. Dynamics of plantation establishment.** The annual evolution of new plantations established per producer region from 1984 onwards.

**The age structure of oil palm plantations in Southeast Asia**
Figure 3 shows the distribution of oil palm plantation age in Southeast Asia in 2017, which differs significantly across the major producing regions. Malaysia has the longest history of large-scale plantation expansion in the region[16]. This is reflected in an average plantation age of around 13.1 years in Insular Malaysia, whereas Kalimantan has the most recent expansion frontier[19], with plantation age averaging 7.4 years. Thailand, with an average of just below eight years, has a similarly young age structure. Sumatra and Peninsular Malaysia have medium-aged plantations of 10 years each on average. For strategic management of the oil palm sector, it is relevant that we detect a very young plantation age structure on Kalimantan where 51% of plantations are aged less than seven years and are therefore still to reach the most productive age. Sumatra has the most balanced age structure with 2.08 Mha (40% of the island's plantation area) being at a prime age, while 27% (1.4 Mha) are old, and the remainder of 1.73 Mha or 33% is still young. In Insular Malaysia, on the other hand, 380,000 hectares, or more than 45% of plantations, have surpassed the most productive stage and will need to be replanted once they have passed 25 years of age.

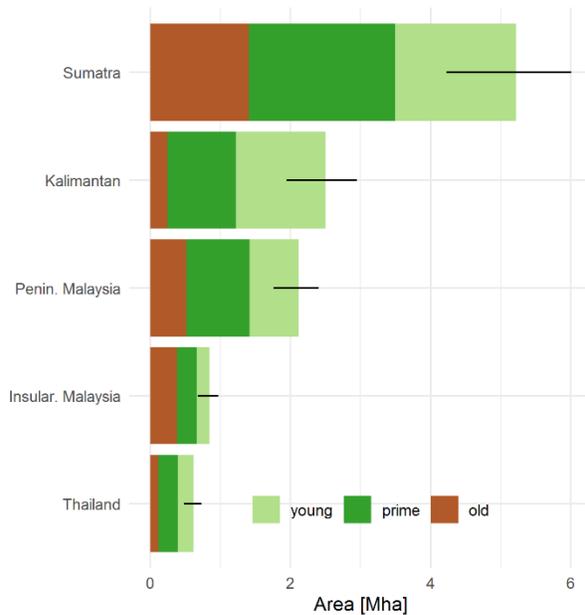

**Figure 3. Age distribution of oil palm plantations.** The age composition of each of the five main oil palm producing regions for functional age classes (young <7 years; prime 7-15; old >15 years). Error bars show the 95% confidence interval of the total oil palm area.

This study has combined oil palm detection with vegetation disturbance detection using free and open high-resolution satellite data along a timeline, thereby assessing the extent and age of oil palm plantations in Southeast Asia. This approach has a number of potential applications that can be scaled up globally or used for national level monitoring in the future as outlined below.

**Identifying location and spatial extent of oil palm plantations**

The proposed methodology can reliably identify oil palm plantations independent of weather, daylight conditions or costly ground-sourced information amidst other tree crops in the landscape. Furthermore, this algorithm is able to detect smallholder plantations with a high spatial resolution. This is particularly valuable in light of recent studies pointing to the important role of smallholders as drivers of plantation expansion[33,34]. The role of smallholder plantations might be accentuated in the future, in Indonesia in particular, by the anticipated zero-deforestation, zero-peat and zero-fire commitments to be applied to industrial plantations[14,35,36], as well as the potential general moratorium on the issuance of new industrial oil palm plantations[37,38]. Sustainability standards usually prescribe provisions in space (i.e., no expansion on forest and peat) and time (i.e., a cut-off date after which forest clearing is prohibited). In this regard, the high spatial and temporal resolution of the oil palm map produced here will allow for refined definition and assessment of compliance with these standards

**Identifying hotspots of change**

Existing legal and regulatory requirements and sustainability standards restrict the conversion of forests to oil palm plantations. In this regard, at least 20% of total global oil palm supply is certified under the Roundtable on Sustainable Palm Oil (RSPO) standard, which prescribes no forest conversion whatsoever[37,38] on land; government-led initiatives such as the Sustainable Palm Oil Standards of Indonesia and Malaysia (ISPO and MSPO, respectively), as well as zero-deforestation pledges made by more than a hundred oil palm sourcing companies[15].

The monitoring of company commitments is often performed by commercial service providers by means of remote sensing. The methodology presented here allows for monitoring clearings and subsequent planting in real-time. The combination of remotely sensed information with spatial information about estate boundaries allows specific actors and their adherence to environmental legislation and standards to be identified.

**Identifying yield and production using spatial extent and age structure**

The age of plantations is an important predictor of oil yield as palm age influences the quality and quantity of the fresh fruit bunches[16]. The remote sensing product presented here can assist high-level planning of the oil palm sector to adapt its strategies regarding plantation management. For example, our results highlight that Kalimantan has a skewed age structure in which young plantations are overrepresented. These plantations will come into full production in the next five years and will significantly increase oil production, thereby relaxing the need for new conversions in the future. In contrast, the majority of plantations in Peninsular Malaysia are at a prime age now, and existing young plantations cannot make up for the anticipated aging and consequent shortfall in oil palm production in the next decade. The sector will require major investments and coordination to rejuvenate these old plantations to establish a more balanced age structure.

**Integrating Earth Observation with national inventories**

Complementing national inventories with the type of products produced here has several benefits: i) it increases transparency as governments or civil society can track agro-forest management; ii) it supports the calculation of spatially-explicit estimates of greenhouse gas emissions and removals, as set out in recent UN guidance on REDD+ measurement, reporting and verification (MRV)[40]; iii) it increases the frequency of assessments, allowing timely policy responses; and iv) it complements official statistics with independent checks. The proposed methodology, using high-resolution imagery under "free, full and open" licenses, can provide the foundations for a dedicated United Nations Framework Convention on Climate Change (UNFCCC) MRV framework for forest management under the Paris Agreement.

In summary, our analysis has shown that Earth Observation can provide timely, independent, transparent and consistent monitoring of oil palm across large geographical areas, complementing official statistics and bridging the gap between technology and land policy. The results reveal a striking and previously undocumented dynamic expansion of the oil palm sector in Southeast Asia over the last three decades. While we acknowledge the economic benefits of oil palm plantations in terms of employment and income generation, their harvest rate should be carefully evaluated to avoid negative impacts on the carbon stock and ecosystem services such as biodiversity, soil protection, and water quality. We believe that timely and robust evidence from space observation, as presented here, are essential tools for assessing these potential trade-offs. Such an approach can also be used for improving the monitoring of oil palm and other crops at a global scale.

## Methods

**Data**

The oil palm maps are the result of time-series analyses of Copernicus (i.e., Sentinel-1 and Sentinel-2) and Landsat (i.e., Landsat 5 and 7) archives characterizing oil palm extent and change with a spatial resolution of ~ 30 m (the spatial resolution varies slightly with latitude and the sensor). For detection of oil palm in particular, the main input data required for this analysis included: 1) the Copernicus Sentinel-1 microwave time series; 2) the Sentinel-2 multi-spectral time series; 3) the Landsat multi-spectral time series; and 4) additional auxiliary data sets.

Copernicus is the European Union's Earth Observation program, which provides global, and continuous Earth Observation capacity. The Sentinel-1 (A and B) and Sentinel-2 (A and B) satellites have been providing images operationally as part of the European Copernicus program since 2014 and 2015, respectively. The Sentinel-1 mission collects data using a dual-polarization C-band Synthetic Aperture Radar (SAR) instrument. Over land, data are acquired in the so-called interferometric wide mode, in swaths of approximately 185 km. The data are available in the Google Earth Engine (GEE) catalogue where the Ground Range Detected (GRD) Level 1 products are pre-processed with the Sentinel-1 Toolbox. The processing steps include thermal noise removal, radiometric calibration and terrain correction, resulting in an application-ready data product that includes the backscattering coefficients in VV (vertically transmitted, vertically received) and VH (vertically transmitted, horizontally received) polarizations with a 10 m pixel spacing. The Sentinel-1 sensors have a 12 day repeat orbit, and since the identical A and B versions are interleaved, this is an effective revisit of six days in either descending (local morning) or ascending (local evening) passes. However, the Sentinel-1 observation scenario limits acquisitions over the study area to a single 12-day descending orbit over Kalimantan, Sarawak and Sabah, and a combination of 12-day ascending and descending orbits over Sumatra, Peninsular Malaysia and Thailand.

The Sentinel-2 mission provides high-resolution multi-spectral images, with each image consisting of 13 spectral bands in the visible/near infrared (VNIR) and shortwave infrared spectral range (SWIR). We derived the Normalized Difference Vegetation Index (NDVI) and additional features from Sentinel-2 data and used them for post-classification filtering of the oil palm map. The Sentinel data have a resolution of 10 m. The Landsat data were obtained as Top of the Atmosphere (TOA) corrected reflectance at a 30 m resolution. Furthermore, we used digital elevation data from the Shuttle Radar Topography Mission (SRTM) at a resolution of one arc-second (approximately 30 m). All the data described above were accessed using the Google Earth Engine[41].

Medium-resolution imagery from the Landsat archives spanning back as far as 1984 was used to estimate the age of plantations in 2017. Landsat 5 and 7 are low Earth orbit satellites launched in 1984 and 1999, respectively. Landsat 5 carries the Thematic Mapper (TM) sensors while Landsat 7 carries the Enhanced Thematic Mapper Plus (ETM+) sensor (an improved version of the TM instruments). Both instruments provide high-resolution multi-spectral images in the visible/near infrared (VNIR) and shortwave infrared spectral range (SWIR) with a spatial resolution of 30 m.

Additional auxiliary data comprise thematic masks and digital elevation data for filtering during post-processing, extraction of the oil palm statistics and the creation of maps for cross-comparison of the oil palm area. Mangroves have a similar spectral signature as oil palms, leading to false classification as oil palms. Therefore, we compiled a mangroves layer to reliably mask out mangrove forest stands. The resulting layer considers the widest possible extent from three global mangrove maps[45–47]. Furthermore, we used digital elevation data from the SRTM at a resolution of one arc-second (approximately 30 m) to define terrain slope.

**Google Earth Engine**
Google Earth Engine (GEE) is a cloud-based platform that enables large-scale geospatial analyses[41]. It includes functions to handle raster and vector data, as well as the necessary implementation tools for analyzing the data and applying machine-learning approaches. GEE hosts a large catalogue of publicly available geospatial data sets, data from the EU Copernicus Sentinel missions, Landsat archives, topographic maps and socioeconomic data. All processing is free and runs in the cloud, meaning that implemented approaches do not rely on local infrastructure. This makes it possible to scale up developed approaches without the need to download the data or upgrade local clusters. The GEE platform was developed for researchers, scientists and developers to assist in the use of satellite and spatial data across different fields. All data classification for this study was performed in GEE because it provides the ability to build a reproducible pipeline for pixel-level classification with high computational efficiency.

**Oil palm area detection**
Oil palm area detection requires preparation of Sentinel-1 data and the application of an unsupervised oil palm detection routine based on NDVI and texture features. The combination of NDVI and texture is important to reliably distinguish oil palm from other woody vegetation. To this end, an annual mosaic of backscatter radar for 2017 was produced by processing all available Sentinel-1 data (1A and 1B) with single polarizations (VV and VH). The main objectives in producing this annual mosaic were: (i) to build a baseline C-band backscatter of Southeast Asia; and (ii) to assess radar backscatter changes between different years in land cover change studies.

The value of the output pixels corresponds to the 'mean' value of the Sentinel-1 collection after: i) outlier removal (eliminating the lowest 20% of values); and ii) normalization for incidence angle to remove some of the ascending/descending variation. In addition to the VV and VH yearly mean, we added VH/VV difference and summed average (SAVG) texture metrics from the Gray Level Co-occurrence Matrix around each pixel of every band (radius of three pixels). To calculate SAVG, all input data were scaled to bytes.

**Land use classification**

We deployed a stratified unsupervised classification algorithm to detect oil palm without training data. The workflow is presented in Figure 4. First, to account for regional differences, we divided Southeast Asia into 12 equally-sized grids with a side length of five degrees and then randomly allocated 50,000 training points within each grid. We used unsupervised classification with yearly means of VV, VH and the VV/VH-ratio derived from Sentinel-1 images as input features and SAVG texture metrics from the Gray-Level Co-Occurrence Matrix of those features. For each grid, the detection algorithm returns the optimal number of clusters (between 10 and 16) to fit the sample. Subsequently, we identified clusters of oil palm areas through the visual interpretation of high-resolution imagery and applied the oil palm classification from each grid to the entire study area, resulting in twelve, at times contradicting, oil palm maps. We reconciled these competing classifications by applying a majority filter, i.e., the final product shows oil palm where seven or more of the 12 classifications identify oil palm.

Post-processing was used to apply rule-based corrections to the intermediate oil palm map resulting from the unsupervised classification. These corrections were necessary in settlements, mangroves and sloping areas, as well as for other objects like roads running through the plantations. From a reflectance viewpoint, settlements are very heterogeneous, such that the backscattered values in some parts of settlements are similar to those observed from oil palm fields. Setting a minimum NDVI value of 0.5 removes settlements from the area classified as oil palm. Similar confusion exists with mangroves. This was resolved by deploying the mangrove mask described in the Data section and removing areas falsely classified as oil palm from within the mangrove areas.

On Kalimantan, coverage by Sentinel satellites is in descending orbit only. This results in gaps in some areas on slopes. To account for these gaps on south sloping areas in Kalimantan, we applied a post-processing majority filter within a 3x3 (30x30m) moving window to fill gaps in detected oil palm plantations. This process introduced a de facto minimum mapping unit of 900 m$^2$ for the oil palm map. For a refinement of the final product, we removed pixels with NDVI less than 0.5, which at times have been erroneously introduced by the majority filtering in small non-palm areas such as roads running through plantations, etc.

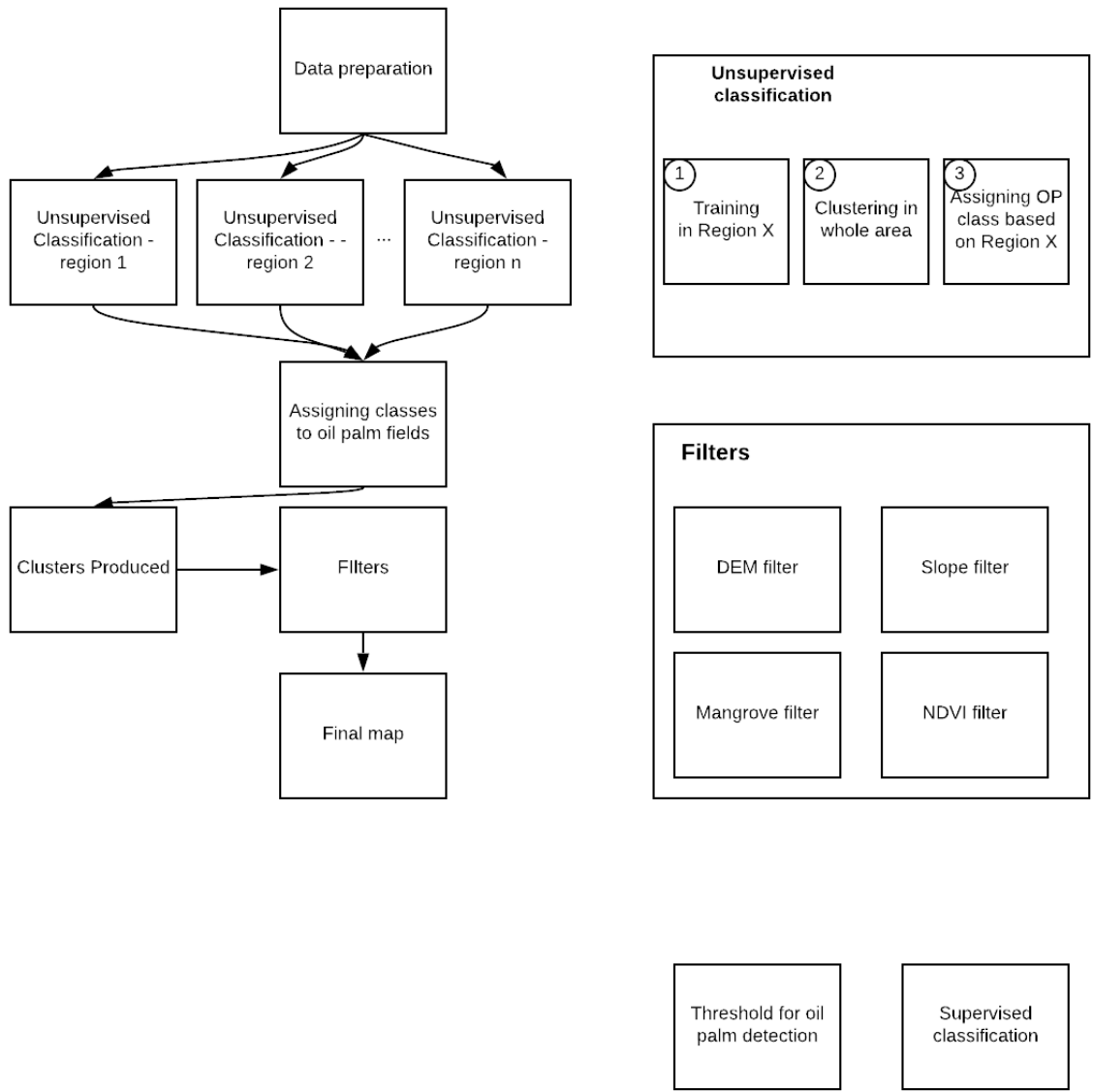

**Figure 4.** The unsupervised Sentinel 1-based oil palm detection routine involving thematic clusters and filtering during post-processing.

**Oil palm age detection**

As an oil palm canopy develops over time, the area fraction of bare soil decreases. This development can be observed from optical remote sensing through the bare soil index (BSI). To estimate plantation age, we moved backwards in time from oil palm plantations detected in 2017 until the BSI exceeded a certain value in young stands. To this end, we used surface reflectance values from the Landsat 5 and Landsat 7 image collections. We masked out clouds for every scene using the pixel quality attributes generated from the CFMASK algorithm[45,46]. We then generated a time series of BSI starting from 1984. To smooth the time series and to remove noisy observations, we calculated the median from a moving window of 12 months of the time series and omitted any 12-month time period in which we had less than three observations.

To determine a reference BSI for oil palm, we calculated a histogram of BSI values from established plantations in 2017 and used the 95th percentile as a threshold value. The 95th percentile coincides with the presence of very young plantations (two to three years of age with an open canopy and a high BSI), as confirmed by the visual interpretation of high-resolution imagery. For all pixels above the threshold in 2017, we analyzed the BSI time series backwards in time where canopy closure is defined as the point when the BSI index drops below the cut-off value. This means that the resulting oil palm age map will register the first observation of oil palm plantations at an age of two to three years.

The resulting oil palm age map has similarities, but also notable differences with the annual Hansen tree loss maps[30]. The Landsat-based approach to detecting vegetation disturbances is common to both. The main difference is the time dimension: our approach defines the status (oil palm plantations) in 2017, then travels back in time through the Landsat archive to detect the first major disturbance in the canopy (a high BSI), representing the last moment before oil palms closed the canopy. The Hansen approach defines the status (tree cover) in the year 2000, then travels forward in time to detect the first clearing. These methodological differences result in remarkable differences in the final product, e.g., the Hansen data set consistently detects disturbances earlier than we do. This is to be expected, as in reality, there is often a gap of five years or more between clearing forest and establishing oil palm[19], in addition to the time needed (two to three years) for the oil palms to close the canopy.

**Map validation**

We determined the accuracy of our oil palm map against a systematically collected and independently interpreted sample of very high-resolution images. We used a stratified random sample design for a total of 40,000 locations in Thailand, Malaysia and Indonesia, which was then visually interpreted using very high-resolution images performed by 44 anonymous volunteers through a citizen science platform (geo-wiki.org/games/picturepile/)[47]. Previous data collection campaigns with Picture Pile have shown high accuracies in identifying cropland[48] and building damage assessment[49], where the majority is used to combine multiple observations at the same location and/or observations are compared with control points interpreted by experts. The map accuracy ranges from 80% for Thailand to 85% for Peninsular Malaysia with no clear trend in commission or omission of reference oil palm plantations.

Table 1. Mapping accuracies assessed using independent validation samples.

| Region | # samples | Balanced Accuracy [%] | 95% CI lower bound [%] | 95% CI upper bound [%] |
|---|---|---|---|---|
| **Kalimantan** | 1,555 | 80.00 | 77.92 | 81.96 |
| **Sumatra** | 840 | 83.06 | 81.02 | 84.96 |
| **Insular Malaysia** | 1,109 | 82.78 | 80.42 | 84.96 |
| **Peninsular Malaysia** | 1,608 | 84.95 | 83.11 | 86.66 |
| **Thailand** | 1,525 | 80.33 | 78.24 | 82.30 |

**Comparison with other studies**

We acquired four other existing spatial data sets[6,18,50,51] reporting the extent of oil palm in Southeast Asia, along with national statistics from Indonesia[52], Malaysia[53] and Thailand[1]. These spatially explicit products differ from our product in scale (two products,[18] focus on one or two regions only), scope (mapping industrial plantations only[5,6]; mapping full oil palm extent, whereas we do not map very young and otherwise open canopies); resolution (on-screen mapping includes big blocks of plantations including all the related infrastructure[54]; relatively lower 100 m pixel resolution[50]; the reference year (ranging from 2013/14[5,19,51] to 2016[50] whereas we map the extent in 2017); or precision (differentiating oil palm from rubber plantations poses problems for ALOS-PALSAR-based products[51]).

It should be noted that for the above-mentioned reasons, area-wise comparison of existing data sets constitutes the proverbial comparison of apples and pears. The present study detects established oil palm plantations with closed canopy, i.e., those that are older than 2-3 years, and not seriously damaged by pests, wind, fire or other calamities. The other studies listed in Figure 5 and visualized online[1], in contrast, are based on contextual information and include, among others, temporarily unstocked or very young (<2-3 years) plantations, or those that are partly or fully damaged by calamities. These studies further include infrastructure such as roads, oil palm mills and landing areas and, at times, settlements located amidst blocks of plantations.

---

[1] https://olhadanylo.users.earthengine.app/view/oilpalmmapscomparison

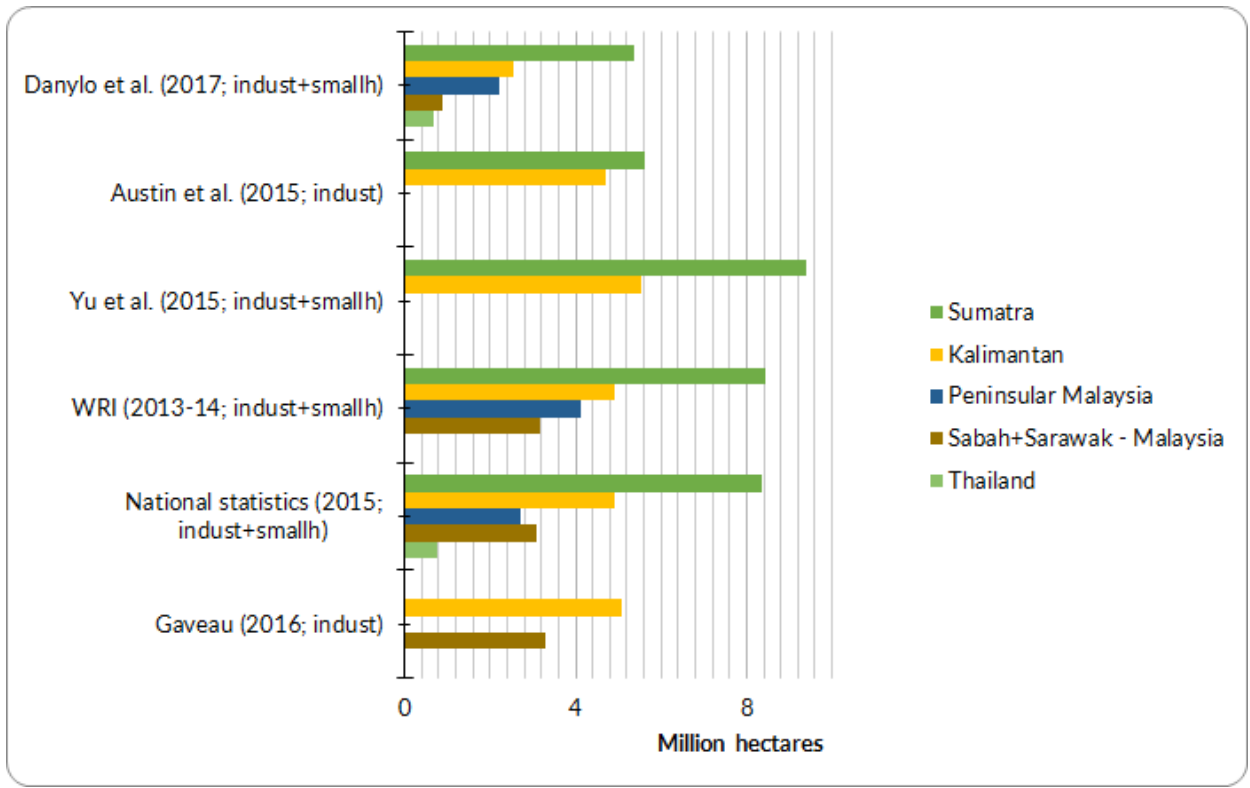

**Figure 5. Comparison of area statistics of our study with other data sets**. The y-axis details the source, the reference year of the maps and the scope (industrial/smallholder plantations). The scale in terms of main oil palm producing regions covered is listed in the legend.

**Known issues and planned improvements**
A few shortcomings of the present methodology and data set are known. First and foremost, since the detection algorithm works on the canopy structure specific to palms, it picks up wild palms and, at times, other riverine woody vegetation if they occur in large tracts, in addition to cultivated oil palm plantations. This is specifically true for mangroves and the stand forming Nypa palm (*Nypa fruticans*), which is native to Southeast Asia but invasive in riparian and estuarine areas in central Africa. In contrast, coconut plantations are not picked up by the detection algorithm, probably due to the wider spacing between trees. Context-specific external thematic data sets can be used to mask out these areas. Plantations stocking on hillsides, particularly on north-facing slopes, are often falsely shown as being patchy. This was corrected using filters during post-processing. Very young plantations (less than 2-3 years of age) could not be detected, as oil palm features at this stage are not separable from annual/perennial crops.

**Acknowledgements**
This work was supported by the RESTORE+ project (www.restoreplus.org) which is part of the International Climate Initiative (IKI), supported by the Federal Ministry for the Environment, Nature Conservation, Building and Nuclear Safety (BMU) based on a decision adopted by the German Bundestag.

**Author Contributions**

OD, JP and GL and GC contributed equally to the work presented in the paper, generating the maps, calculating the statistics and writing the paper. LS, H, IC, FK and SF provided intellectual inputs to the development of the oil palm map on the technical side and on the narrative of the paper. All authors provided edits and suggestions to the manuscript.

**Code Availability**

A web interface powered by Google Earth Engine allows the expert system to be run on any Sentinel1 and Landsat 5 and 7images. Access can be provided on request.

**Data Availability**

The Sentinel and Landsat imagery used in this study is available in Google Earth Engine: https://olhadanylo.users.earthengine.app/view/danylopalmoilagefin.